\documentclass{article}
\usepackage{cite}
\usepackage{graphicx}
\usepackage{dcolumn}

\begin{document}

\date{}
\title{Two-dimensional hydrogen-like atom in a uniform magnetic field. Large-order
perturbation theory}
\author{Francisco M. Fern\'{a}ndez \thanks{%
E-mail: fernande@quimica.unlp.edu.ar} \\
INIFTA, Divisi\'on Qu\'imica Te\'orica\\
Blvd. 113 S/N, Sucursal 4, Casilla de Correo 16, 1900 La Plata, Argentina}
\maketitle

\begin{abstract}
The well known hypervirial perturbation method (HPM)\ based on hypervirial
relations and the Hellmann-Feynman theorem is suitable for the calculation
of perturbation corrections of large order for the two-dimensional
hydrogen-like atom in a uniform magnetic field. We show analytical results
in terms of the quantum numbers and large order corrections for particular
states. This approach appears to be simpler and more efficient than the
recently proposed one based on Green functions.
\end{abstract}

\section{Introduction}

\label{sec:intro}

The two-dimensional hydrogen-like atom in a uniform magnetic field
perpendicular to the atomic plane has received considerable attention as a
model for excitons in thin materials such as nano-scale multilayer
semiconductor systems. The Schr\"{o}dinger equation is separable in polar
coordinates and has been solved approximately in many different ways\cite
{AH67,ST70,MR86,ESM89,ZCX90,MRM92,HG93,RR03,HPL13,HHL13,FILU15,HNHL16,S18}.
Here we are interested in the application of perturbation theory that
provides approximate solutions in terms of power series of the field
strength that are suitable in the weak-field limit\cite
{MR86,RR03,HHL13,FILU15,HNHL16,S18}.

Several authors have obtained perturbation corrections of low order\cite
{MR86,RR03,HHL13,FILU15,S18} but the most extended calculation was due to
Szmytkowski\cite{S18} who solved the perturbation equations of first and
second order by means of Green functions. His analytical results, which are
valid for all quantum numbers, suggest that some of the earlier results\cite
{HHL13,FILU15,HNHL16} may be wrong. The Green-function method is a powerful
approach that requires considerable ingenuity and mastery of the technique;
however, it appears to be unsuitable for the calculation of perturbation
corrections of larger order because it soon becomes rather intractable.

There are other strategies for the calculation of the perturbation
corrections for simple quantum-mechanical models that are more efficient\cite
{F00} and the purpose of this paper is to discuss one of them. Taking into
account the interest in the two-dimensional hydrogen-like atom in a magnetic
field and the discrepancy between earlier perturbation calculations, present
analysis appears to be fully justified.

In section~\ref{sec:model} we outline the model and derive a dimensionless
eigenvalue equation for the radial part of the Schr\"{o}dinger equation that
is suitable for the application of perturbation theory. In section~\ref
{sec:HPM} we briefly develop the chosen perturbation technique and apply it
to the model mentioned above. Finally, in section~\ref{sec:conclusions} we
summarize the main results of the paper and draw conclusions.

\section{Model}

\label{sec:model}

The model consists of a two-dimensional hydrogen-like atom in a
uniform magnetic field perpendicular to the atomic plane. It may
also represent the effective-mass equation for an electron-hole
pair in the presence of a uniform magnetic field\cite{AH67,ST70}.
For concreteness, in what follows we resort to the notation of
Szmytkowski\cite{S18} in order to facilitate comparison of the
results. In the case of a symmetric gauge the
Schr\"{o}dinger equation is separable in polar coordinates ($0\leq r<\infty $%
, $0\leq \phi <2\pi $). If we write the eigenfunctions as
\begin{equation}
\psi (r,\phi )=\frac{1}{\sqrt{r}}P(r)\frac{1}{\sqrt{2\pi }}e^{im_{l}\phi
},\;m_{l}=0,\pm 1,\pm 2,\ldots ,  \label{eq:psi}
\end{equation}
the radial part of the Schr\"{o}dinger equation becomes\cite{S18}
\begin{equation}
\left[ -\frac{\hbar ^{2}}{2m}\frac{d^{2}}{dr^{2}}+\frac{\hbar ^{2}\left(
l^{2}-1/4\right) }{2mr^{2}}+\frac{e\hbar B}{2m}m_{l}-\frac{Ze^{2}}{4\pi
\epsilon _{0}r}+\frac{e^{2}B^{2}r^{2}}{8m}\right] P(r)=EP(r),
\label{eq:Schrod}
\end{equation}
where $m$ and $-e$ are the electron mass and charge, respectively, $Ze$ is
the charge of the spinless nucleus (clamped at the coordinate origin), $%
l=|m_{l}|$ and $B$ is the field intensity. The bound-state solutions satisfy
\begin{equation}
\lim\limits_{r\rightarrow 0}P(r)=0,\;\lim\limits_{r\rightarrow \infty
}P(r)=0.
\end{equation}

Straightforward application of perturbation theory to equation (\ref
{eq:Schrod}) is cumbersome because one has to carry all the physical
constants and model parameters through the whole algebraic process\cite{S18}%
. For this reason it is preferable to convert equation (\ref{eq:Schrod})
into a dimensionless eigenvalue equation. To this end we define a
dimensionless coordinate $q=r/L$, where
\begin{equation}
L=\frac{a_{0}}{Z},\;a_{0}=\frac{4\pi \epsilon _{0}\hbar ^{2}}{me^{2}},
\label{eq:L}
\end{equation}
and equation (\ref{eq:Schrod}) becomes
\begin{eqnarray}
HP(r) &=&\epsilon P(r),  \nonumber \\
H &=&\left[ -\frac{1}{2}\frac{d^{2}}{dq^{2}}+\frac{\xi }{2q^{2}}-\frac{1}{q}%
+\lambda q^{2}\right] ,  \label{eq:Schrod_dim}
\end{eqnarray}
where
\begin{eqnarray}
\epsilon &=&\frac{mL^{2}}{\hbar ^{2}}\left( E-\frac{e\hbar B}{2m}%
m_{l}\right) ,  \nonumber \\
\xi &=&l^{2}-\frac{1}{4},\;\lambda =\frac{e^{2}B^{2}L^{2}}{8\hbar ^{2}}.
\label{eq:lambda,epsilon}
\end{eqnarray}
The unit of energy is
\begin{equation}
\frac{\hbar ^{2}}{2mL^{2}}=\frac{Ze^{2}}{4\pi \epsilon _{0}L}=\frac{%
Z^{2}e^{2}}{4\pi \epsilon _{0}a_{0}},  \label{eq:energy_unit}
\end{equation}
and
\begin{equation}
\frac{e\hbar B}{2m}\frac{mL^{2}}{\hbar ^{2}}=\sqrt{2\lambda }.
\end{equation}
If we define
\begin{equation}
B_{0}=\frac{\hbar }{ea_{0}},  \label{eq:B0}
\end{equation}
then
\begin{equation}
\lambda =\frac{1}{8Z^{4}}\left( \frac{B}{B_{0}}\right) ^{2},
\label{eq:lambda(B/B0)}
\end{equation}
enables us to compare present results with those of Szmytkowski\cite{S18}.
By means of perturbation theory we easily obtain the coefficients of the
perturbation series
\begin{equation}
\epsilon (\lambda )=\sum_{p=0}^{\infty }\epsilon _{p}\lambda ^{p}.
\label{eq:weak-field-series}
\end{equation}

When $B=0$ ($\lambda =0$) the problem is exactly solvable and the allowed
energies are given by\cite{S18}
\begin{equation}
\epsilon _{0}=-\frac{1}{2\left( n_{r}+l+1/2\right) ^{2}},
\label{eq:epsilon_0}
\end{equation}
where $n_{r}=0,1,\ldots $ is the radial quantum number (number of zeroes of $%
P(r)$). Following previous papers we also define $n=n_{r}+l+1$
that resembles the principal quantum number of the
three-dimensional hydrogen atom.

The weak-field series (\ref{eq:weak-field-series}) is divergent. If we
define the new variable $u=\lambda ^{1/4}q$ then the dimensionless operator $%
H$ becomes
\begin{equation}
H=\sqrt{\lambda }\left( -\frac{1}{2}\frac{d^{2}}{du^{2}}+\frac{\xi }{2u^{2}}-%
\frac{\lambda ^{-1/4}}{u}+u^{2}\right) ,  \label{eq:H(u)}
\end{equation}
that clearly shows that we can also expand the eigenvalues in the
strong-field series
\begin{equation}
\epsilon =\sqrt{\lambda }\sum_{j=0}^{\infty }e_{j}\lambda ^{-j/4},
\label{eq:strong-field}
\end{equation}
which has a finite radius of convergence.

\section{Hypervirial perturbation method}

\label{sec:HPM}

One of the most efficient implementations of perturbation theory for
separable problems is the hypervirial perturbation method (HPM) that
combines the hypervirial relations and the Hellmann-Feynman theorem. If $O$
is a linear operator then
\begin{equation}
\left\langle P\right| [H,O]\left| P\right\rangle =0,
\label{eq:hypervirial_gen}
\end{equation}
is called a hypervirial relation, where $[H,O]=HO-OH$ is the commutator
between such pair of operators\cite{F00}. This equation holds if $O$ is
chosen so that $\left\langle P\right| \left. HOP\right\rangle =\left\langle
HP\right| \left. OP\right\rangle $. Another important ingredient of the
method is the Hellmann-Feynman theorem that states that\cite{F00}
\begin{equation}
\frac{\partial \epsilon }{\partial \lambda }=\left\langle \frac{\partial H}{%
\partial \lambda }\right\rangle .  \label{eq:Hell-Feyn}
\end{equation}

If
\begin{equation}
H-\frac{1}{2}\frac{d^{2}}{dq^{2}}+\frac{\xi }{2q^{2}}-\frac{1}{q}+\lambda
q^{K},\;K=1,2,\ldots ,  \label{eq:H_K}
\end{equation}
and $O=\frac{j+1}{2}q^{j}-q^{j+1}\frac{d}{dq}$\cite{F00} then equation (\ref
{eq:hypervirial_gen}) becomes
\begin{eqnarray}
&&2j\epsilon Q_{j-1}+(j-1)\left[ \frac{j(j-2)}{4}-\xi \right]
Q_{j-3}+(2j-1)Q_{j-2}-(2j+K)\lambda Q_{j+K-1}=0,  \nonumber \\
&&Q_{j}=\left\langle q^{j}\right\rangle ,\;j=1,2,\ldots .
\label{eq:hypervirial_Q_j}
\end{eqnarray}
On expanding the expectation values in a Taylor series
\begin{equation}
Q_{j}=\sum_{p=0}^{\infty }Q_{j,p}\lambda ^{p},  \label{eq:Qj_series}
\end{equation}
we obtain
\begin{eqnarray}
Q_{j,i} &=&\frac{1}{2(j+1)\epsilon _{0}}\left\{ j\left[ \xi -\frac{j^{2}-1}{4%
}\right] Q_{j-2,i}-(2j+1)Q_{j-1,i}\right.   \nonumber \\
&&\left. -2(j+1)\sum_{m=1}^{i}\epsilon
_{m}Q_{j,i-m}+(2j+K+2)Q_{j+K,i-1}\right\} ,\;  \nonumber \\
j &=&1,2,\ldots .  \label{eq:Q_j,i}
\end{eqnarray}
In addition to this equation we also have
\begin{equation}
Q_{-1,i}=-2\epsilon _{i}+(K+2)Q_{K,i-1},  \label{eq:Q_-1,i}
\end{equation}
that comes from (\ref{eq:hypervirial_Q_j}) with $j=1$. It is assumed that $%
Q_{0}=1$, so that $Q_{0,i}=\delta _{i0}$ is a starting point for the
recurrences.

In order to obtain the perturbation coefficients $\epsilon _{i}$ and $%
Q_{j,i} $ we need an additional equation provided by the Hellmann-Feynman
theorem (\ref{eq:Hell-Feyn})
\begin{equation}
\epsilon _{i}=\frac{1}{i}Q_{K,i-1},\;i=1,2,\ldots .  \label{eq:epsilon_i}
\end{equation}

The calculation is straightforward and for $K=2$ (present case) we obtain
\begin{equation}
\epsilon _{{1}}=\frac{\left( 2\,n-1\right) ^{2}}{8}\,\left(
-3\,l^{2}+3+5\,n^{2}-5\,n\right) ,
\end{equation}
\begin{equation}
\epsilon _{{2}}=-{\frac{\left( 2\,n-1\right) ^{6}}{1024}}\,\left(
-21\,l^{4}-138\,l^{2}+159-90\,l^{2}n^{2}+90\,l^{2}n+582\,n^{2}-439\,n+143%
\,n^{4}-286\,n^{3}\right) ,
\end{equation}
\begin{eqnarray}
\epsilon _{{3}} &=&{\frac{\left( 2\,n-1\right) ^{10}}{65536}}\,\left(
17967-65495\,n+68835\,n^{4}-107070\,n^{3}+115970\,n^{2}-18066\,l^{2}-35130%
\,l^{2}n^{2}\right.  \nonumber \\
&&\left.
+29910\,l^{2}n+231\,l^{4}-132\,l^{6}-18360\,n^{5}+6120\,n^{6}+10440%
\,l^{2}n^{3}-5220\,l^{2}n^{4}\right) ,
\end{eqnarray}
\begin{eqnarray}
\epsilon _{{4}} &=&-{\frac{\left( 2\,n-1\right) ^{14}}{16777216}}\,\left(
15522195-67825511\,n-4005\,l^{8}+153888490\,n^{4}-180523168\,n^{3}+145662172%
\,n^{2}\right.  \nonumber \\
&&-17506020\,l^{2}-64292340\,l^{2}n^{2}+43194060\,l^{2}n+1991850\,l^{4}-4020%
\,l^{6}-80558702\,n^{5}+33863592\,n^{6}  \nonumber \\
&&+43836660\,l^{2}n^{3}-26018580\,l^{2}n^{4}-1640100\,l^{2}n^{6}+4920300%
\,l^{2}n^{5}-502740\,l^{4}n^{3}+2563680\,l^{4}n^{2}  \nonumber \\
&&\left.
-2312310\,l^{4}n-3060\,l^{6}n^{2}+3060\,l^{6}n+251370\,l^{4}n^{4}-6009164%
\,n^{7}+1502291\,n^{8}\right) ,
\end{eqnarray}
The first three perturbation corrections agree with those of Szmytkowski\cite
{S18} to which we added $\epsilon _{4}$ that was not reported before as far
as we know. We can obtain as many perturbation corrections as desired by
means of the recurrence relations (\ref{eq:Q_j,i}), (\ref{eq:Q_-1,i}) and (%
\ref{eq:epsilon_i}) and available computer algebra software.
However, the perturbation corrections in terms of the quantum
numbers may probably be too long for any practical use. If one is
interested in weak-field expansions of large order it appears to
be preferable to obtain them for particular values
of $n$ and $l$ be means of those recurrence relations. For example, when $%
n=1 $ and $l=0$ we have

\begin{eqnarray}
\epsilon _{{1}} &=&3/8,\;\epsilon _{{2}}=-{\frac{159}{1024},\;}\epsilon _{{3}%
}={\frac{17967}{65536},\;}\epsilon _{{4}}=-{\frac{15522195}{16777216},\;}%
\epsilon _{{5}}={\frac{5189052801}{1073741824}}  \nonumber \\
\epsilon _{{6}} &=&-{\frac{4896676641339}{137438953472},\;}\epsilon _{{7}}={%
\frac{3094900497137871}{8796093022208},\;}\epsilon _{{8}}=-{\frac{%
20233178231139761499}{4503599627370496},}  \nonumber \\
\epsilon _{{9}} &=&{\frac{20808558827825859998445}{288230376151711744},\;}%
\epsilon _{{10}}=-{\frac{52693485465369543566065089}{36893488147419103232},\;%
}  \nonumber \\
\epsilon _{{11}} &=&{\frac{80639435078901048406195920633}{%
2361183241434822606848},\;}\epsilon _{{12}}=-{\frac{%
587353055515797037508553136130823}{604462909807314587353088},}  \nonumber \\
\epsilon _{{13}} &=&{\frac{1255613239147236284205667622925365349}{%
38685626227668133590597632}}  \nonumber \\
\epsilon _{{14}} &=&-{\frac{6229668057619980010555555519950165544755}{%
4951760157141521099596496896}}  \nonumber \\
\epsilon _{{15}} &=&{\frac{17753264589549239693872523415436400485638255}{%
316912650057057350374175801344}}  \nonumber \\
\epsilon _{{16}} &=&-{\frac{921721759137179716887942948086717222595277533675%
}{324518553658426726783156020576256}}  \nonumber \\
\epsilon _{{17}} &=&{\frac{%
3379056665253674076167201632469154672196055608756005}{%
20769187434139310514121985316880384}}  \nonumber \\
\epsilon _{{18}} &=&-{\frac{%
27797116247667972439940810526714208588100705850127986405}{%
2658455991569831745807614120560689152}}  \nonumber \\
\epsilon _{{19}} &=&{\frac{%
127484555261829518463134910686385252583016203699835125715445}{%
170141183460469231731687303715884105728}}  \nonumber \\
\epsilon _{{20}} &=&-{\frac{%
2593203450314371618931792865686398116783507010792581025252777725}{%
43556142965880123323311949751266331066368}}
\end{eqnarray}

It is worth noting that the HPM cannot be applied to the operator (\ref
{eq:H(u)}) for the calculation of the strong-field series. The reason is
that the unperturbed operator $H_{0}=\lim\limits_{\lambda \rightarrow \infty
}\lambda ^{-1/2}H$ is an even function of the coordinate $u$ and therefore
the recurrence relations do not yield the expectation values $%
U_{j}=\left\langle u^{j}\right\rangle $ for odd and even $j$ necessary for
the application of the technique and the calculation of $U_{-1}$.

\section{Conclusions}

\label{sec:conclusions}

The calculation of perturbation corrections of large order by means of
Greens functions does not appear to be practicable. On the other hand, the
HPM yields simple straightforward recurrence relations for the systematic
calculation of the perturbation corrections to the energy and expectation
values of powers of the coordinate. These recurrence relations are suitable
for programming in available computer algebra systems which greatly
facilitates the systematic calculation of those perturbation corrections to
any desired order.

Present results agree with those of Szmytkowski\cite{S18} and not
with those derived by other authors by means of the Levi-Civita
transformation of the coordinates and the expression of the
resulting two-dimensional Hamiltonian operator in terms of
suitable creation-annihilation operators\cite
{HHL13,FILU15,HNHL16}. We have no doubt that their weak-field
expansions do not agree with the actual ones. For example, the
weak-field expansions derived by Hoang et al\cite{HNHL16} are not
correct because they come from approximate variational expressions
for the energies. However, it is worth mentioning that for
sufficiently large field intensities they fit the actual
eigenvalues more accurately than the exact weak-field expansions
because the latter are divergent.

\end{document}